\documentclass[draft,grl]{agutex}

\usepackage{graphicx,color}

\authorrunninghead{LINCOT ET AL.}

\titlerunninghead{Inner-core anisotropy from hcp alloy}

\usepackage{multirow}

\usepackage{graphicx}
\begin{document}

%
%

%
%

\title{Multiscale model of global inner-core anisotropy induced by hcp-alloy plasticity}

\authors{A. Lincot\altaffilmark{1,2},  Ph. Cardin\altaffilmark{1}, R. Deguen\altaffilmark{3}, S. Merkel\altaffilmark{2,4}}

\altaffiltext{1}{Universit\'e Grenoble Alpes, CNRS, ISTerre, F-38041 Grenoble, France.}

\altaffiltext{2}{UMET, Unit\'e Mat\'eriaux et Transformations, CNRS, INRA, ENSCL, Universit\'e de Lille, F-59000 Lille, France.}

\altaffiltext{3}{Laboratoire de g\'eologie de Lyon, ENS de Lyon, Universit\'e Lyon-1, F-69007 Lyon, France.}

\altaffiltext{4}{Institut Universitaire de France, F-75005 Paris, France}

%
%


\begin{abstract}
The Earth's solid inner-core exhibits a global seismic anisotropy of several percents. It results from a coherent alignment of anisotropic Fe-alloy crystals through the inner-core history that can be sampled by present-day seismic observations. By combining self-consistent polycrystal plasticity, inner-core formation models, Monte-Carlo search for elastic moduli, and simulations of seismic measurements, we introduce a multiscale model that can reproduce a global seismic anisotropy of several percents aligned with the Earth's rotation axis. Conditions for a successful model are an hexagonal-close-packed structure for the inner-core Fe-alloy, plastic deformation by pyramidal $\langle c+a \rangle$ slip, and large-scale flow induced by a low-degree inner-core formation model. For global anisotropies ranging between 1 and 3\%, the elastic anisotropy in the single crystal ranges from 5 to 20\% with larger velocities along the $c$-axis. 
\end{abstract}

\keypoints{
 \item Multiscale model of inner-core anisotropy produced by hcp alloy deformation
 \item 5 to 20\% single-crystal elastic anisotropy and plastic deformation by pyramidal slip
 \item Low-degree inner-core formation model with faster crystallization at the equator
}

%
%

\begin{article}

\section{Introduction}

Seismic observations using body waves differential travel times \citep{poupinet1983,morelli1986grl,waszek2011ng}, long period normal modes \citep{woodhouse1986grl,deuss2010s}, and the analysis of autocorrelation of earthquake coda
\citep{wang2015ng} provide strong evidences that the Earth's inner-core is anisotropic, with $P$-waves traveling faster by up to 3\% in the polar than in the equatorial direction. Further analyses refined this observation, providing evidences for both hemispherical and radial variations of the amplitude of anisotropy \citep{tkalcic2015revgeophys,deuss2014areps,wang2015ng}. However, this apparent complexity should not obscure the first-order observation that the fast propagation direction for inner-core seismic waves is aligned with the Earth's rotation axis. The observed axial anisotropy still lacks a conclusive explanation. This paper will hence focus on reproducing this first-order observation from a multiscale model. 

\begin{figure}
\centering
\includegraphics[width=14cm]{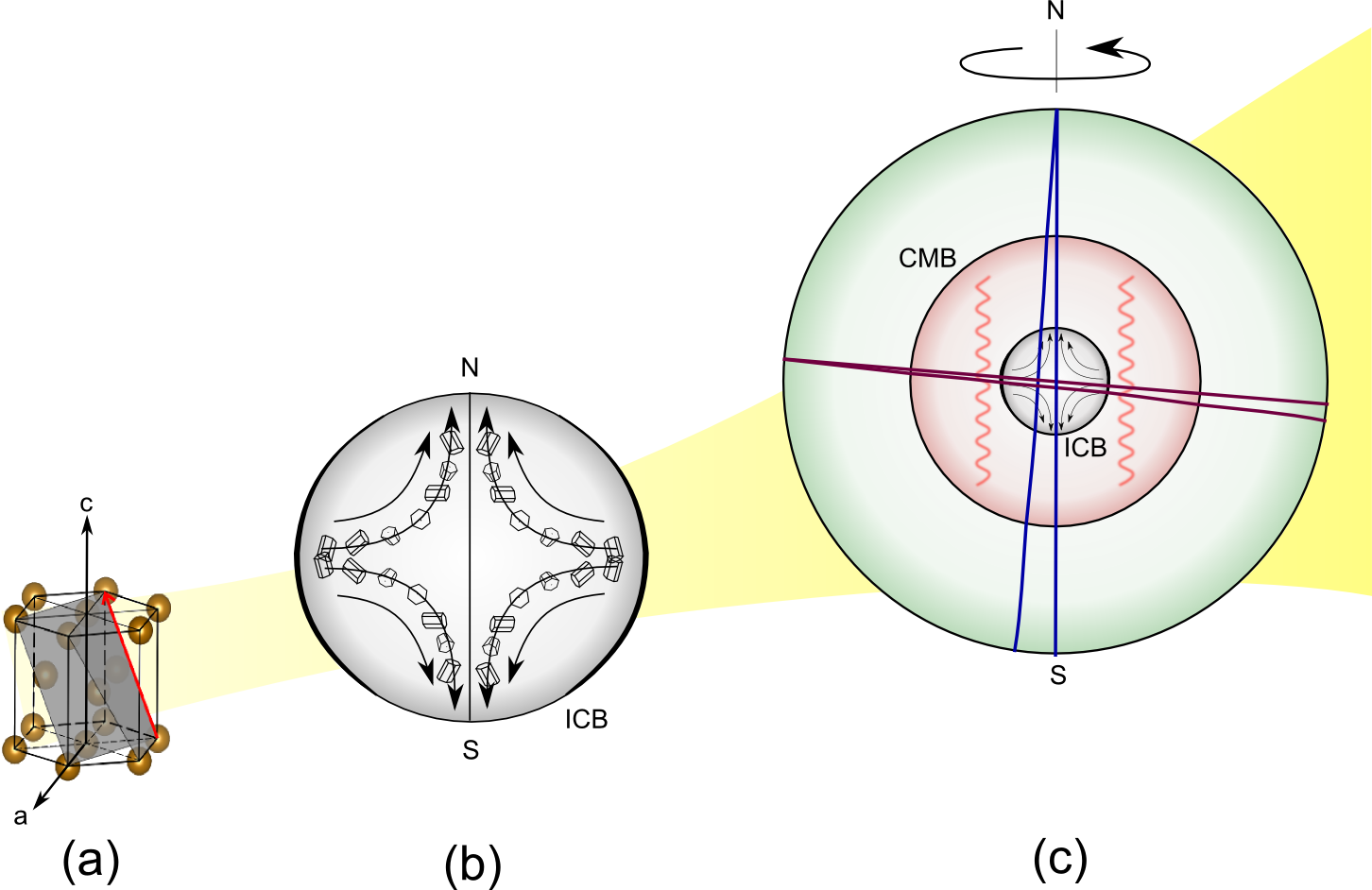}
\caption{\label{fig:multiscale}Multi-scale model of inner-core anisotropy. Anisotropic hexagonal Fe-alloy crystals \textit{(a)} plastically rotate through slip under the action of deformation in the inner-core \textit{(b)}. Deformation in the inner-core depends on the core formation model, driven by heat extracted by outer core convection \textit{(c)}. At the global scale, seismic anisotropy is measured with body-wave differential travel times  \textit{(c)} which are faster along the North-South axis (blue trajectories) than in the equatorial plane (red trajectories).
}
\end{figure}

A multiscale model of inner-core anisotropy should allow to build a synthetic inner-core with a given choice of crystal structure, elastic moduli, and crystal alignment mechanism, coupled with an inner-core formation hypothesis. In a second stage, the model should simulate present-day seismic observations which can then be compared with actual measurements \citep[Fig.~\ref{fig:multiscale},][]{lincot2014crg}.   

In this work, we investigate whether such integrated model can indeed reproduce the first-order observation of several percents North-South global seismic anisotropy in the inner-core. The abundance of symmetries for elasticity and plasticity of body-centered-cubic (bcc) and face-centered-cubic (fcc) phases combined with the integrated nature of inner-core anisotropy measurements is such that plastic deformation of such crystal structures cannot explain the global inner-core anisotropy \citep{lincot2015grl}.
Here, we investigate the effect of an hexagonal-close-packed (hcp) structure for the inner-core Fe alloy. Under the assumption that the inner-core anisotropy results from plastic deformation along a dominant slip system, we find that necessary conditions for a successful anisotropic model are 5 to 20\% single crystal elastic anisotropy, plastic deformation by pyramidal slip, and a large-scale flow induced by a low-degree inner-core formation model.

\section{Methods}

\subsection{Inner-core formation model}

It is typically assumed that the inner-core anisotropic structure results from an alignment of anisotropic Fe-alloy crystals acquired either during solidification \citep{bergman1997,deuss2010s} or as a result of subsequent deformation \citep{yoshida1996,karato1999,wenk2000jgr,deguen2012epsl} although other models have been proposed assuming, for instance, the presence of aligned aspherical liquid inclusions in the inner-core \citep{singhSC2000}.

Radial dendritic growth \citep{bergman1997} aligns the crystals relative to the spherical inner-core boundary (ICB) which is unlikely to produce a  global cylindrical axisymmetric anisotropy \citep{lincot2014crg}. An additional orientation mechanism is required for aligning the fast $P$-wave crystal directions with the Earth's rotation axis. Two classes of processes have been suggested in the literature: \textit{(i)} a forcing of inner-core dynamics from the outer core magnetic field \citep{karato1999,buffet2001} \textit{(ii)}  an effect of the Earth's rotation through the ellipticity of a convective inner-core \citep{buffett2009gji} or inner-core preferential growth \citep{yoshida1996}.  Recent studies evaluated that the direct effect of the Lorentz forcing generates weak deformation \citep{lasbleis2015gji}. The effect of the magnetic field on anisotropy would result from solidification textures mostly, which are not well understood at inner-core conditions and have not been observed in analogue experiments 
\citep{brito2002pepi}. In addition, recent calculations of the thermal conductivity of Fe 
at core conditions \citep{zhang2015nature} indicate that thermal convection is unlikely in the inner-core \citep{deguen2012epsl}, although compositional convection may be an option \citep{gubbins2013grl}.

\begin{figure}
\begin{center}
\includegraphics[width=14cm]{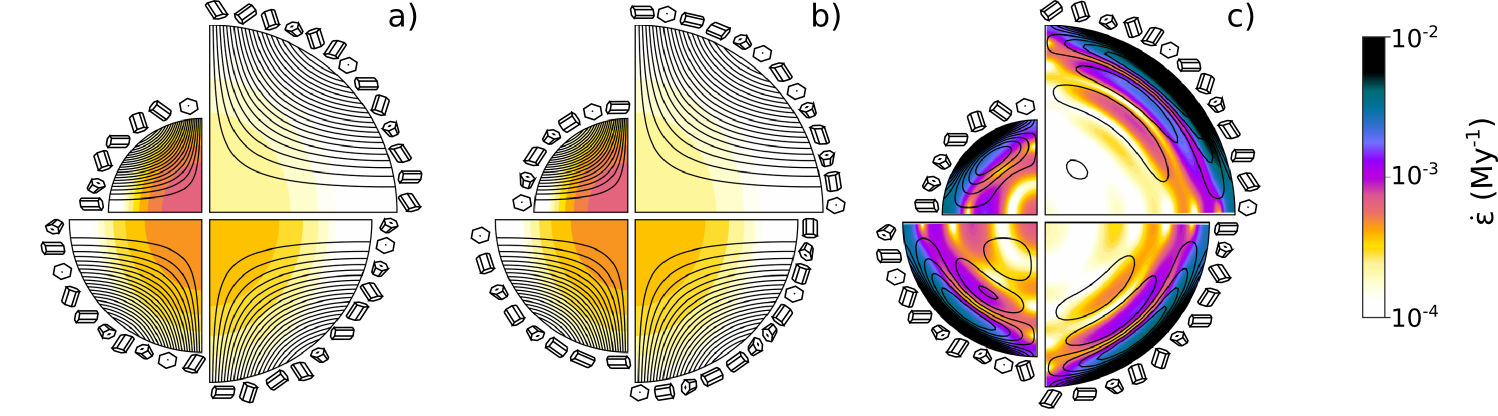} \vspace*{0.5cm}
\end{center}
\caption{\label{fig:models}
Inner-core formation models with preferential growth of the inner-core at the equator \citep{yoshida1996}. The resulting topography is continuously relaxed by a quadrupolar flow, generating plastic deformation of the inner core Fe-Alloy. 
Model \textit{a)} assumes random crystallization textures at the ICB while model \textit{b)} adds solidification textures in which the $c$-axes of the hcp crystals lie preferentially in the plane of the ICB. 
Model \textit{c)} accounts for a stable density stratification during inner-core formation. Each figure shows the time evolution of flow during the inner-core history at non-dimensional times t = 0.25, 0.5, 0.75 and 1, where 0 is the formation of the inner-core and 1 is the present day, starting anticlockwise from the upper left quadrant. Colors are Von-mises equivalent strain rate $\dot{\epsilon}$ (in Myr$^{-1}$, log scale). Black lines in overlay are representative streamlines.}
\end{figure}

We therefore focus on models of inner-core growth with faster solidification in the equatorial plane \citep[Fig.~\ref{fig:models}a,][]{yoshida1996}, with solidification texturing \citep[Fig.~\ref{fig:models}b,][]{bergman1997} and/or density stratification \citep[Fig.~\ref{fig:models}c,][]{deguen2009ng} as possible additional ingredients. In all cases, we impose a  solidification rate twice as large at the equator than at the poles. Model \textit{a)} assumes random solidification textures while model \textit{b)}  includes solidification textures in which the $c$-axes of the hcp crystals lie preferentially in the plane of the ICB.  Model \textit{c)} accounts for a stable density stratification during inner-core formation. The importance of stratification depends on the value of the buoyancy number $\mathcal{B}^{*}=\frac{\Delta \rho\, g\, r_{ic}^{2}}{\eta\, \dot r_{ic}}$, where $\Delta \rho$ is the density difference across the inner core associated with compositional or 
thermal stratification (excluding the contribution of compression), $g$ the acceleration of gravity at the ICB, $\eta$ the dynamic viscosity of the inner core, $r_{ic}$ the inner core radius and $\dot r_{ic}$ the inner core growth rate \citep{deguen2009ng}.
Density stratification tends to localize the deformation in a thin shear layer below the ICB, whose thickness decreases with increasing $|\mathcal{B}|$, and in which the strain rate increases with $|\mathcal{B}|$.
In this study, we set $\mathcal{B}^{*}=- 10^{6}$  (corresponding to $\Delta \rho \simeq 5$ kg.m$^{-3}$  and  $\eta\simeq 10^{18}$ Pa.s). Larger $\mathcal{B}^{*}$ values would result in earlier stratification in the Earth's history and and, consequently, reduce the central part of pure shear deformation \citep{deguen2009ng,deguen2011pepi}. Also note that model \textit{c)} assumes random solidification textures at the ICB. Deformation is large in such model and, hence, solidification textures are quickly erased and do not play an important role in the modeled global scale anisotropy, as in this study, but can play a crucial role for studies focused on the superficial layer of the inner-core.

As illustrated in Fig.~\ref{fig:models}, a model reproducing present day seismic observations should integrate the whole history of the inner-core. In that view, seismic anisotropy is a marker of inner-core formation and evolution and will record changes in dynamics or mineralogy during its history.

\subsection{Inner-core Fe-alloy textures}

\begin{figure}
\centering
\includegraphics[width=14cm]{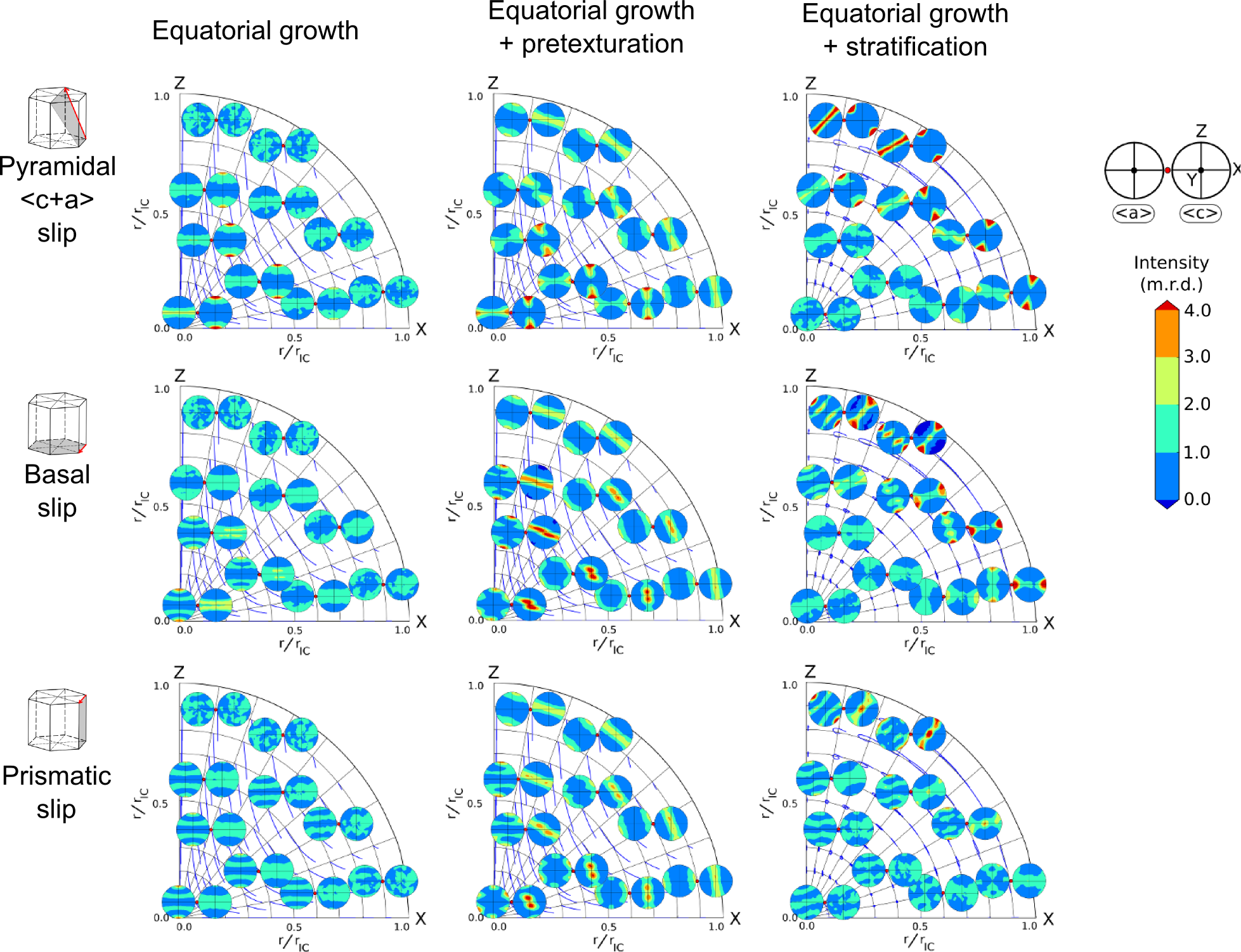}
\caption{\label{fig:textures}Pairs of pole figures of the $a$ (left) and $c$ (right) directions representing present-day textures of the inner-core hcp alloy. Calculated with dominant pyramidal $\langle c+a \rangle$ slip (top row), basal slip (middle row), or prismatic slip (bottom row). Core formation models with fast crystallization at the equator (left), with the addition of  crystallization textures at the ICB (center), or density stratification (right). The vertical $Z$ axis is the geographical South-North axis while $X$ lies in the equatorial plane. Blue lines are the trajectories of the polycrystalline aggregates  after crystallization at the  ICB. Contours in multiples of a random distribution (m.r.d., linear scale). $r_{ic}$ is the inner-core radius. Pyramidal slip produces the strongest global textures, with the $c$-axes of the aggregate mostly aligned with the Earth's rotation axis. Simulation with other slip systems produce less consistent textures, in which the $c$-axes of the aggregates can be 
found in multiple 
orientations.}
\end{figure}

For each inner-core formation model, we assume plastic deformation of the inner-core Fe-alloy on a dominant slip system. We follow the deformation of 100 markers introduced at the ICB during inner-core growth and calculate the corresponding textures for a 10000 grains aggregate using the visco-plastic self-consistent (VPSC) code of \cite{lebensohn1993AMM}.  Experiments on Fe and analogues indicate that basal slip and twinning dominate the plastic behavior of hcp-Fe at low temperature  \citep{wenk2000nature,merkel2004pepi} and that, as temperature increases, twinning disappears and the activity of pyramidal $\langle c+a \rangle$ slip increases \citep{poirier2002,miyagi2008jap,merkel2012msmse}. We simulate the texture of an hcp aggregate at each marker for three different dominant slip systems: basal, prismatic and pyramidal $\langle c+a \rangle$ (Table \ref{table:vpsc}). The final results are meridional maps of textures for a present day inner-core (Fig.~\ref{fig:textures}).

This results in a striking observation: simulations with the pyramidal $\langle c+a \rangle$ slip system are very efficient at aligning the $c$-axes of the polycrystals with the Earth's rotation axis. Simulations with other slip systems can induce strong textures at the local scale but, over the scale of the inner core, the hcp-alloy alignment is not consistent. Indeed, unlike basal and prismatic slip, pyramidal $\langle c+a \rangle$ slip includes $c$ in the slip direction: it is prone to align the $c$-axes of the aggregate during deformation. 

\subsection{Inner-core elasticity}

Given the spread of published data for the elasticity of hcp-Fe at inner-core conditions \citep{vocadlo2009epsl,sha2010grl,martorell2013science} and the unknown influence of alloying elements, we estimate the global inner-core anisotropy as a function of  single-crystal elasticity over a wide range of values. Rather than using grid points in the five-dimensional space of independent elastic constants, we generate elastic moduli by a Monte-Carlo approach. 

We generate 4500  random sets of single-crystal elastic moduli which satisfy the following requirements. The Hill average of the bulk and shear moduli of the randomly oriented aggregate should be within 15\% of inner-core values \citep[$K=1400$~GPa and $G=170$~GPa,][]{PREM}. The single-crystal elastic moduli should also satisfy conditions for mechanical stability for hcp \citep{wallace1972} 
\begin{eqnarray}
& &  C_{11}-C_{12} > 0 ; \\
& & C_{11}+C_{12}+C_{33} > 0 ; \nonumber \\
& & (C_{11}+C_{12})C_{33}-2\;C_{13}^2 > 0 ; \nonumber \\
& & C_{44} > 0, \nonumber
\end{eqnarray}
where $C_{ij}$ are single-crystal elastic moduli. 

For each inner-core formation model, elastic model, and marker position,  we calculate the polycrystalline elastic tensor using the simulated textures and single-crystal elastic moduli.

\subsection{Inner-core seismic anisotropy}

More than $300000$ synthetic seismic rays are generated randomly to probe the whole inner-core. For each ray, we estimate the normalized seismic travel times residual
$\delta{t}/{t}  = (s - s^{0})/{s^{0}}$
where $s$ is the simulated slowness of the seismic ray, 
and $s^0$ is the slowness of that same ray for an homogeneous and fully
isotropic inner-core \citep{lincot2014crg}.

Many seismological studies use averaging procedures with travel times residuals fitted to \citep{creager1992nature,deuss2014areps}
\begin{equation} 
 \delta{t}/{t} = a_1 + a_2\cos^{2}\zeta + a_3\cos^{4}\zeta,
 \label{eqAniso}
\end{equation}
where $a_1$, $a_2$ and $a_3$ are adjustable parameters and $\zeta$ is the angle
between the ray and the Earth rotation axis. The quantity $a_2 + a_3$ is the
difference between polar ($\zeta=0^\circ$) and equatorial ($\zeta=90^\circ$) residuals. 
It is a measure of the global inner-core anisotropy often reported in the
literature. It will be used in the rest of the paper (Fig.~\ref{fig:anisotropy}).

\section{Results}

\subsection{Parameters for hcp single-crystal anisotropy}

Despite the complexity introduced by elasticity, crystal plasticity, geodynamics, and seismic rays geometry,
two dimensionless elastic parameters are found to primarily affect the global inner-core anisotropy:  $\Delta^{c-a}$ and $\Delta^{45^\circ}$. $\Delta^{c-a}$ describes the relative difference 
in $P$-waves velocities along the $c$- and $a$-axes ($V_P^{c}$ and $V_P^{a}$), and $\Delta^{45^\circ}$ is related to the concavity of the single-crystal $P$-wave velocities, $45^\circ$ away from the $c-$axis ($V_P^{45^\circ}$). They are calculated as follow
\begin{eqnarray}
  \Delta^{c-a}  & = &  \frac{V_P^{c} -
V_P^{a}}{V_P^{m}}, \label{eqAn1} \\
  \Delta^{45^\circ}  & = &  \frac{V_P^{45^\circ} -
\left( V_P^{c} + V_P^{a} \right)/ 2}{V_P^{m}}, \label{eqAnisMono} \nonumber
\end{eqnarray}
where $V_P^{m}$ is the average $P$-wave velocity. 

Global anisotropy results are consistent when plotted as a function of $\Delta^{c-a}$ and $\Delta^{45^\circ}$: they reveal parallel isocontour mainly controlled by $\Delta^{c-a}$ (Fig.~\ref{fig:anisotropy}). This analysis allows to classify our 4500 sets of random elastic moduli and other calculations from the literature in a simple representation and estimate their effect on global inner-core anisotropy.

\subsection{Inner-core anisotropy}

Seismic studies report global inner core anisotropy of up to several percents with North-South velocities larger than those in the equatorial plane  \citep{deuss2014areps,tkalcic2015revgeophys}. As shown in Fig.~\ref{fig:anisotropy}, this level of anisotropy is difficult to reach if deformation of the inner-core hcp-alloy is controlled by dominant basal or prismatic slip, irrespectively of the inner-core formation model. Indeed, inner-core anisotropy is measured with waves crossing the entire inner core. If polycrystalline textures are not consistent over large regions, the anisotropy of the single-crystal will be canceled at the global scale of the seismic measurement.

Pyramidal $\langle c+a \rangle$ slip, on the other hand, is prone to align the $c$-axes of an aggregate during its deformation. Hexagonal crystals are axisymmetric around the single-crystal $c$-axis for elasticity. As such, this mechanism allows for significant enhancement of the global scale inner-core anisotropy. With dominant pyramidal $\langle c+a \rangle$ slip, faster velocities along single-crystal $c$-axes ($\Delta^{c-a} > 0$) are required  to obtain a positive global anisotropy. $\Delta^{45^\circ}$ acts as a secondary but non negligible factor: a concave variation of the velocities ($\Delta^{45^\circ}>0$) tends to amplify the global anisotropy.

Models with dominant pyramidal $\langle c+a \rangle$ slip, faster crystallization in the equatorial region, and random crystallization textures at the solidification front produce global anisotropies 5 to 15 times smaller than $\Delta^{c-a}$ at the single-crystal level. The addition of solidification textures enhances the global anisotropy but also adds scatter in the representation of Fig.~\ref{fig:anisotropy} with a few models with $\Delta^{c-a} <10\%$ producing a global anisotropy larger than 3\%. In such model, textures produced in the central portions of the inner-core are induced by plastic deformation, with a global cylindrical symmetry, whereas the outer portions of the inner-core are mostly influenced by solidification textures with a global spherical symmetry (Fig.~\ref{fig:textures}). This induces heterogeneities in travel time calculations and scatter in the representation of Fig.~\ref{fig:anisotropy}.
The addition of stratification reinforces the global anisotropy by up to 50\%. For example, in the stratified model with dominant pyramidal $\langle c+a \rangle$ slip, single-crystal anisotropies with $\Delta^{c-a}$ between 5 and $20\%$ are required for a global anisotropy between 1 and $3\%$. 

\section{Discussion and conclusion}

Previous studies have shown that bcc and fcc phases orientation through dislocation creep cannot explain the observed North-South global inner-core anisotropy \citep{lincot2015grl} . Here, we show that an hcp inner-core Fe-alloy, deforming plastically along a dominant pyramidal $\langle c+a \rangle$ slip system, with 5 to 20\% elastic anisotropy in the single-crystal, and a simple low-degree inner-core formation model can produce a global anisotropy of 1 to 3\%.

Hexagonal crystals are axisymmetric around their $c$-axis for elasticity. As such, a most efficient model of inner-core anisotropy should align the $c$-axes of the hcp aggregates with the Earth's rotation axis. This ingredient is crucial for reproducing inner-core anisotropy and can be obtained if the plasticity of the inner-core hcp-alloy is dominated by pyramidal $\langle c+a \rangle$ slip. 

\begin{figure}
\centering
\includegraphics[width=14cm]{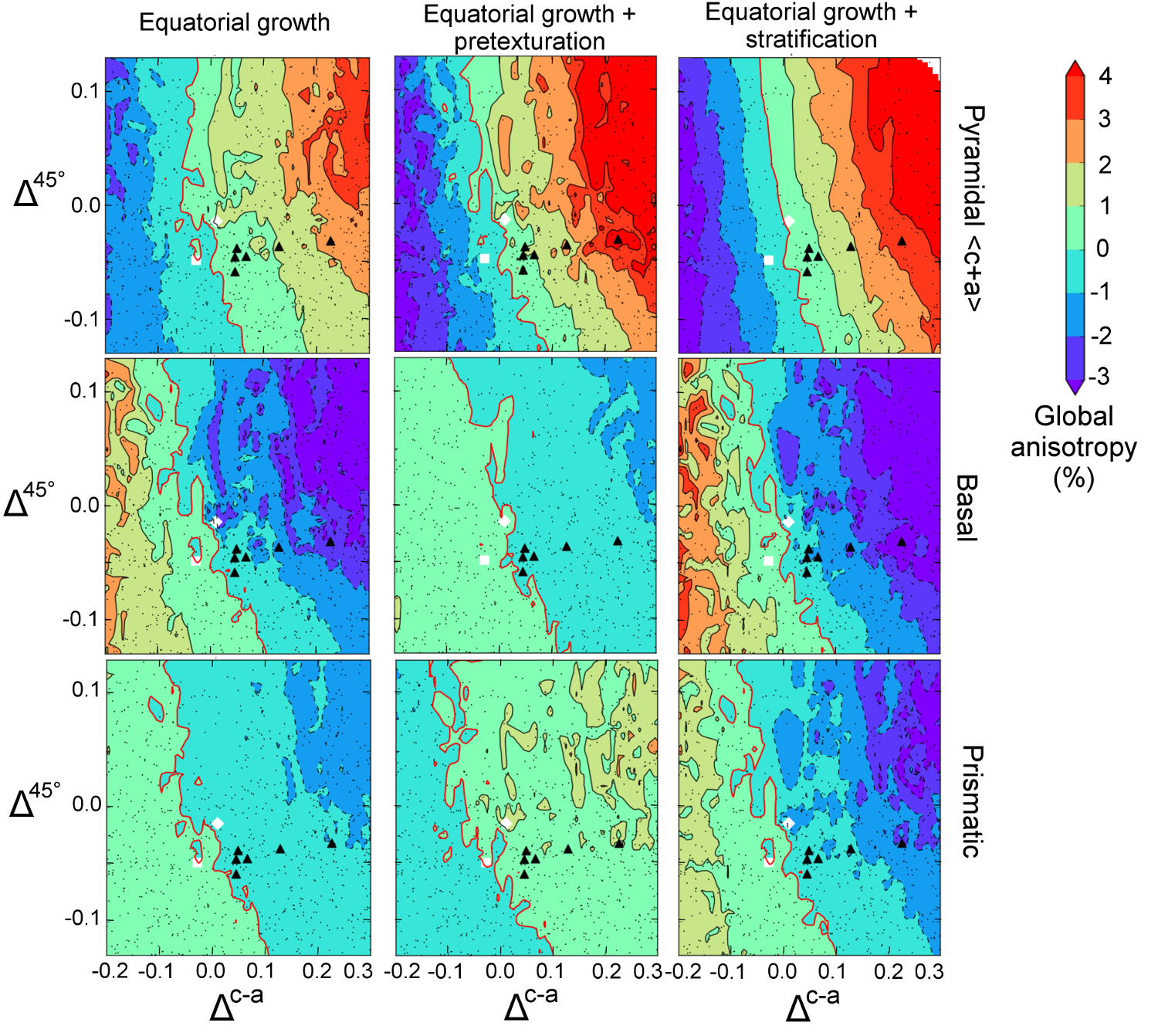}
\caption{\label{fig:anisotropy}Global inner-core anisotropy (color scale) vs. single-crystal anisotropy parameters $\Delta^{c-a}$ and $\Delta^{45^\circ}$ for different core-formation models (columns) and dominant hcp slip systems (rows). Calculations were performed for 4500 random sets of elastic moduli  (black dots) and the contour lines are obtained by interpolation.
First-principles calculations for hcp-Fe single-crystal elasticity at inner-core conditions are shown as white diamond for \cite{vocadlo2009epsl}, white square for \cite{sha2010grl}, and black triangles for \cite{martorell2013science} that include an effect of premelting.}
\end{figure}

Fig.~\ref{fig:anisotropy} allows for constraining  $\Delta^{c-a}$ and $\Delta^{45^\circ}$ for single-crystal elastic tensors compatible with anisotropy observations. Do note, however, that our solution for single-crystal elasticity is not unique.
First-principles calculations for stable hcp-Fe at inner-core conditions \citep{vocadlo2009epsl,sha2010grl} lead to a weak single-crystal anisotropy and weak global anisotropy, below 1\% (Fig.~\ref{fig:anisotropy}). Recently, it was shown that hcp-Fe may exhibit a strong nonlinear shear weakening just before melting, with a reduction in $V_S$ and a large increase in single-crystal anisotropy \citep{martorell2013science}. With such elastic moduli, our multi-scale model is successful at reproducing inner-core global seismic anisotropy (Fig.~\ref{fig:anisotropy}).

A large-scale coherent alignment of anisotropic Fe-alloy through the inner-core is required to reproduce a North-South global seismic anisotropy, which likely requires that crystal alignment is due to a low-degree deformation field. A more complex geometry like, for example, high Rayleigh number convection, could generate locally strong textures but such anisotropy would be canceled at the scale of seismic observations. 

At this point, other details of the scenario of inner-core formation can not be  discriminated. 
Observations of inner-core anisotropy are constrained by a distinct set of specific ray paths -- a few hundreds -- that are now being supplemented by the addition of virtual paths using the cross-correlation of ambient noise and earthquake coda \citep{wang2015ng,boue2014epsl}. In addition, the representation of seismic observations with fits (Eq.~\ref{eqAniso}) oversimplifies the data by smoothing the geographical and depth dependence of the seismic travel times. In the future, such direct multi-scale model should be extended for  inverting the actual  observations of seismic travel times, including their radial and lateral variations \citep{deuss2014areps,tkalcic2015revgeophys,wang2015ng}. 
This new approach will constrain the different scenarios regarding the history of the Earth's inner-core and its  interactions with the outer core \citep{deguen2012epsl} and the geodynamo  \citep{aubert2008Nat}.

\begin{acknowledgments}
The authors wish to thank Mathieu Dumberry, the Geodynamo group at ISTerre, and the Mineral Physics group at UMET for useful discussions as well as Mike Bergman and Shigeo Yoshida for constructive reviews. This work has been financed by the program PNP of CNRS/INSU and labex OSUG@2020. RD acknowledges support from grant ANR-12-PDOC-0015-01 of the ANR (Agence Nationale de la Recherche). Calculations were made at Centre de Calcul Commun of the OSUG. Data used in this paper are available upon request.
\end{acknowledgments}

\bibliographystyle{agu}

\newpage
\clearpage

\begin{table}
\begin{center}
\begin{tabular*}{\hsize}{@{\extracolsep{\fill}}llcccc}
\hline
\multirow{2}{*}{Slip system} & & \multicolumn{4}{c}{Dominant slip system} \\ 
  &&Basal& Prismatic & Pyr. $\langle c+a \rangle$  1& Pyr. $\langle c+a \rangle$ 2\\
\hline 
Basal & $\left(0001\right)\langle 11\overline{2}0 \rangle$ & 0.5 & 1.0 & 2.0 & 2.0 \\
Prismatic & $\left\{1\overline{1}00\right\}\langle 11\overline{2}0\rangle$ & 1.0 & 0.5 & 2.0 & 2.0 \\
Pyr. $\langle a \rangle$ & $\left\{1\overline{1}01\right\}\langle 11\overline{2}0\rangle$ & 3.0 & 3.0 & 3.0 & 3.0 \\
Pyr. $\langle c+a \rangle$ 1st order & $\left\{\overline{1}011\right\}\langle 11\overline{2}3\rangle$ & 2.0 & 2.0 & 0.5 & $\infty$ \\
Pyr. $\langle c+a \rangle$ 2d order & $\left\{\overline{1}\overline{1}22\right\}\langle 11\overline{2}3\rangle$ & $\infty$ & $\infty$ & $\infty$ & 0.5 \\
\hline 
\end{tabular*}
\end{center}
\caption{\label{table:vpsc}Normalized critical resolved shear stresses used for calculating the development of texture in the hcp aggregate for models with the following dominant slip systems: basal, prismatic, pyramidal $\langle c+a \rangle$ first order, pyramidal $\langle c+a \rangle$ second order. The pyramidal $\langle a \rangle$ slip system is weakly activated in all cases. Results for simulations using both pyramidal $\langle c+a \rangle$ slip systems are nearly identical and are presented together in the paper.}
\end{table}

\clearpage
\cleardoublepage

\end{article}
\end{document}